\documentclass[10pt,aps,prc,showpacs,nofootinbib,superscriptaddress,preprintnumbers,
twocolumn]{revtex4-1}

\usepackage[utf8]{inputenc}
\usepackage{graphicx,amsmath,amssymb,bbm}
\usepackage[dvipsnames]{xcolor}
\usepackage{lmodern,dsfont}
\usepackage[normalem]{ulem}  
\usepackage{hyperref}
\hypersetup{colorlinks=true,urlcolor=blue,linkcolor=blue,citecolor=blue}

\newcommand{\barr}[1]{\bar{\bar{#1}}}
\newcommand{\be}{\begin{eqnarray}}
\newcommand{\ee}{\end{eqnarray}}
\newcommand{\non}{\nonumber \\}

\begin{document}

\title{Three-body Unitarity in the Finite Volume}

\author{M.\ Mai}
\email{maximmai@gwu.edu}
\affiliation{
The George Washington University,
 Washington, DC 20052, USA}

\author{M.\ D\"oring}
\email{doring@gwu.edu}
\affiliation{
The George Washington University,
 Washington, DC 20052, USA}
\affiliation{Thomas Jefferson National Accelerator Facility, Newport News, VA
23606, USA}

\date{\today}
\preprint{JLAB-THY-17-2554}
\begin{abstract}
The physical interpretation of lattice QCD simulations, performed in a small volume, requires an extrapolation to the infinite volume.
A method is proposed to perform such an extrapolation for three interacting particles at energies above threshold. For this, a recently formulated relativistic $3\to 3$ amplitude based on the isobar formulation is adapted to the finite volume. The guiding principle is two- and three-body unitarity that imposes the imaginary parts of the amplitude in the infinite volume. In turn, these imaginary parts dictate the leading power-law finite-volume effects. It is demonstrated that finite-volume poles arising from the singular interaction, from the external two-body sub-amplitudes, and from the disconnected topology cancel exactly leaving only the genuine three-body eigenvalues. The corresponding quantization condition is derived for the case of three identical scalar-isoscalar particles and its numerical implementation is demonstrated.

\end{abstract}


\maketitle

\section{Introduction}

Three-body dynamics plays a major and sometimes dominant role in the understanding of hadronic resonances. In the meson sector~\cite{Guo:2017jvc}, new excited states are searched for in large campaigns such as the new GlueX experiment at Jefferson Lab~\cite{AlGhoul:2017nbp}, the COMPASS experiment~\cite{Alekseev:2009aa}, and BESIII~\cite{Asner:2008nq}. Finding an exotic state with quantum numbers that cannot only be composed of two constituent quarks would provide evidence for the need of explicit gluon dynamics in the description of mesons. Furthermore, many exotic but also conventional mesons decay dominantly or even exclusively into three particles such as the $a_1(1260)$~\cite{Patrignani:2016xqp}. 

In the baryon sector, explored at CLAS/JLab~\cite{Ripani:2002ss}, ELSA~\cite{Sokhoyan:2015fra}, MAMI~\cite{Ahrens:2003na}, and other facilities the $\pi\pi N$ channels provide a substantial source of inelasticity and become dominant at higher energies. But even at low energies they can be crucial as in case of the Roper resonance $N(1440)1/2^+$ that, despite its low mass, has large branching ratios into the $\pi\pi N$ channels. 

The aim of this work is to provide an amplitude to analyze three-body dynamics in lattice QCD simulations. Scattering amplitudes exhibit a continuous spectral function above threshold. In contrast, simulations of QCD in a cube with periodic boundary conditions produce a discrete spectrum of energy eigenvalues. Such finite-volume effects are large above threshold and they become even more relevant as quark masses come closer to their physical values 
and bound states become resonances. Yet, if the interaction region is well confined within the cube, finite-volume effects can in fact be used to determine exactly one scattering phase shift for one energy eigenvalue of the QCD Hamiltonian through L\"uscher's method~\cite{Luscher:1986pf, Luscher:1990ux}. Subsequently, a chiral extrapolation to physical quark masses, see, e.g.,~\cite{Bolton:2015psa, Hu:2017wli, Doring:2016bdr, Guo:2016zep, Hu:2016shf, Guo:2016zos}, provides physical two-body phase shifts.

If more than one two-body channel is on-shell (e.g., the $\pi\pi/K\bar K$ system at energies above the $K\bar K$ threshold), there are still methods to extrapolate the scattering amplitude to the infinite volume~\cite{Hu:2017wli, Briceno:2017max, Hu:2016shf, Doring:2016bdr, Guo:2016zep, Guo:2016zos, Liu:2016wxq, Wu:2016ixr, Agadjanov:2016fbd, Sasaki:2015ifa, Hall:2014uca, Molina:2015uqp, MartinezTorres:2012yi,  Briceno:2015tza, Briceno:2015csa, Briceno:2014uqa, Briceno:2014oea, Torres:2014vna, Wu:2014vma, Doring:2013glu, Li:2014wga, Briceno:2012yi, Hansen:2012tf, Guo:2012hv, Doring:2012eu, Doring:2011vk, Doring:2011nd, Doring:2011ip, Bernard:2010fp, Lage:2009zv, Liu:2005kr, He:2005ey}. Excited mesons in coupled channels have been recently determined by the HadronSpectrum collaboration for a variety of quantum numbers~\cite{Briceno:2017qmb, Briceno:2016mjc, Moir:2016srx, Wilson:2015dqa}.
One of the problems for, e.g., two coupled channels $a$ and $b$ is the need to know three channel transitions at a fixed total energy ($a\to a$, $a\to b$, $b\to b$). However, only one energy eigenvalue is usually available at that energy. Therefore, minimal assumptions on the energy dependence have to be made to relate different eigenvalues and to obtain a well-constrained solution. 

If three particles can be on-shell this problem of underdetermination becomes even more involved because there are eight independent kinematic variables in $3\to 3$ scattering~(see, e.g., Ref.~\cite{Mai:2017vot}). In view of these problems we propose a new method for the infinite-volume extrapolation of three-body systems. 
The method should (a) be energy dependent. This allows also to directly extract resonance properties in the infinite volume by analytic continuation to the resonance poles~\cite{Lage:2009zv, Doring:2011vk,Briceno:2017qmb, Briceno:2016mjc, Wilson:2015dqa}. The problem also suggests to perform a partial-wave expansion respecting cubic symmetry that reduces the degrees of freedom, at the cost of having to cut the expansion in a practical calculation. Both approximations can be improved as lattice QCD data improve: through the introduction of a less constrained energy dependence and the inclusion of more partial waves, respectively.

In order to dispose over definite 3-body quantum numbers that allow for a partial-wave decomposition, one can first construct the quantum number of two particles (``isobar'') and then include the third one, called ``spectator''. A more precise definition of isobar is given below, but it should be noted that an isobar in this definition can have but is not required to have resonant behavior in the corresponding two-particle sub-amplitude. 

In addition, the method should (b) respect unitarity in the infinite volume. In the three-body case, this requirement is technically more involved than for two-body (coupled-channel) unitarity. As shown in Ref.~\cite{Mai:2017vot} it is indeed possible to formulate an amplitude that manifestly fulfills three-body unitarity and two-body unitarity for the sub-amplitudes even for energies above the three-body threshold. 
In turn, one can use the constraints from unitarity to identify those parts of the finite-volume amplitude for which all three particles can be on-shell and for which the finite-volume effects are of power-law nature. In other words, the method should (c) provide all finite-volume divergences in agreement with the three-body quantization condition. 

In summary, based on Ref.~\cite{Mai:2017vot} we present a relativistic, energy-dependent three-body amplitude that is manifestly two and three-body unitary in the infinite volume and fulfills the three-body quantization condition in the finite volume. Special attention will be paid to the cancellation of finite-volume divergences associated with two-body scattering that can potentially obscure the genuine three-body dynamics.

In recent years, significant theoretical progress towards the understanding of three particles in the finite volume has been made~\cite{Guo:2017crd, Sharpe:2017jej, Hammer:2017kms, Hammer:2017uqm, Hansen:2017mnd, Guo:2017ism, Guo:2016fgl, Agadjanov:2016mao, Hansen:2016ync, Hansen:2016fzj, Jansen:2015lha, Hansen:2015zta, Hansen:2015zga, Meissner:2014dea, Hansen:2014eka, Briceno:2014tqa, Aoki:2013cra, Briceno:2012rv, Bour:2012hn, Kreuzer:2012sr, Roca:2012rx, Polejaeva:2012ut, Kreuzer:2010ti, Kreuzer:2009jp, Kreuzer:2008bi} including for $2\to 3$ coupled systems~\cite{Briceno:2017tce}. Most of these works concentrate on the explicit parametrization of the three-body amplitude --like the present one-- although in Refs.~\cite{Agadjanov:2016mao, Hansen:2017mnd} methods are proposed to obtain essential information on the system without the need of explicitly parametrizing the full dynamics.

It is instructive to compare the present approach to the recent work of Ref.~\cite{Hammer:2017kms} that is related in the following sense: Instead of formulating a 
L\"uscher-like approach in which the energy eigenvalues are directly related to the scattering matrix (or phase shift in the two-body case), Ref.~\cite{Hammer:2017kms} and the present study propose a two-step process: In a first step, the free internal parameters of a finite-volume amplitude are fitted to energy eigenvalues from a lattice QCD simulation. In a second step, the infinite-volume version of that amplitude is simply evaluated. For the two-body case, this workflow was first demonstrated in Refs.~\cite{Lage:2009zv, Doring:2011vk} and shown to be equivalent to the two-body L\"uscher equation, up to exponentially suppressed contributions. 
Yet, the proposed method differs from the one of Ref.~\cite{Hammer:2017kms} in a few aspects: it is (a) relativistic which has practical consequences for the construction of the finite-volume boosted two-body sub-amplitude (``dimer'' in the language of Ref.~\cite{Hammer:2017kms}). (b) The construction principle here is three-body unitarity and no use of effective field theory is made; the present three-body amplitude can be expressed in terms of on-shell two-body amplitudes and three-body forces~\cite{Mai:2017vot}. (c) The projection of the singular isobar-spectator interaction to the $A_1^+$ irreducible representation (the example chosen here) is performed by summation over lattice points instead of performing a partial-wave integral.

In Refs.~\cite{Wu:2017qve, Wu:2016ixr, Liu:2016uzk} an effective Hamiltonian model for the $\pi N$ system is constructed and $\pi\pi N$ channels are approximated via a stable $\sigma$ meson and a stable $\Delta(1232)$ baryon. This leads to problems with the three-body quantization condition because the finite-volume three-body singularities are qualitatively different from two-body ones and in principle not related at all. 
See also the discussion in Ref.~\cite{Briceno:2012rv} about the problem with the quantization condition encountered in Ref.~\cite{Guo:2013qla}.

As for the perspectives of the proposed method, 
not many data concerning three-body scattering systems exist on the lattice, so far, but rapid progress is being made. With the possibility of three hadrons being on-shell, the use of meson and baryon-like operators to extract the energy eigenvalues, especially of scattering levels, is needed, as, e.g., demonstrated in Ref.~\cite{Lang:2016hnn}. Along these lines, 
pioneering simulations have been carried out for the quantum numbers of the $a_1(1260)$ and $b_1(1235)$ mesons~\cite{Lang:2014tia}. 
In the meson-baryon sector, energy eigenvalues above the $\pi\pi N$ threshold in the $J^P=1/2^+$ sector have been calculated recently~\cite{Lang:2016hnn}, again with the use of up to several two-hadron operators (the $\sigma$ quantum numbers were realized by a local two-quark operator). Employing three hadron operators of the $\pi$, $\pi$, $N$ type, or $\pi$, $\pi$, $\pi$ type for mesons has not yet been realized. References~\cite{Lang:2016hnn, Kiratidis:2016hda} shed new light on the Roper puzzle, i.e., the problem that the Roper resonance could not be found in many lattice QCD simulations; an analysis of the eigenvalues adapting the proposed formalism to a coupled-channel $\pi N$-$\pi\pi N$ system with full three-body dynamics provides an interesting perspective.

This work is organized as follows: In Sec.~\ref{sec:threeinf} the main properties of the infinite-volume three-body amplitude derived in Ref.~\cite{Mai:2017vot} are discussed. In Sec.~\ref{sec:finvol} the finite-volume amplitude is defined. In Sec.~\ref{sec:numeric} the general workflow, the cancellation of singularities, and a numerical implementation are demonstrated.


\section{Three-particle scattering amplitude}
\label{sec:threeinf}

In previous work \cite{Mai:2017vot} we have addressed the form of a relativistic, infinite volume three-body scattering amplitude in the isobar formalism. It has been shown that it can be expressed in terms of on-shell, two-body unitary $2\to 2$ amplitudes plus genuine three-body interactions which are forced to be real by three-body unitarity. The ``isobar'' notation refers to the parametrization of the $2\to 2$ amplitudes in terms of a dressed $s$-channel propagator with dissociation vertices attached to both ends. As further discussed in Ref.~\cite{Mai:2017vot}, the isobar can be associated with bound states, one or more resonances, or a non-resonant two-particle amplitude.  The isobar formulation is not an approximation but a re-parametrization of the full two-body amplitude as shown in Ref.~\cite{Bedaque:1999vb} and also discussed in Ref.~\cite{Hammer:2017kms}. In the following we collect only the main results of the derivation, relevant for this work and refer the reader for details to the original work~\cite{Mai:2017vot}.

The interaction of three spin-less particles of mass $M$ and out- and in-going four-momenta $q_1,q_2,q_3$ and $p_1,p_2,p_3$, respectively, is fully described by the S-matrix ($\mathcal{S}$) related to the T-matrix ($\mathcal{T}$) via ${\mathcal{S}{}=:\mathbbm{1}+i(2\pi)^4 \delta^4\!\left(\sum_{i=1}^3(q_{i}-p_{i})\right) \mathcal{T}}$. In the case of $3\to 3$ scattering the latter consists of a fully connected ($\mathcal{T}_c$) and a once disconnected piece ($\mathcal{T}_d$), related to the isobar-spectator scattering amplitude $T$ and isobar-propagator $\tau$ as
\begin{widetext}
\begin{align}
\label{eq:t33full}
\langle q_1,q_2,q_3|&\mathcal{T}| p_1,p_2,p_3\rangle
=
\langle q_1,q_2,q_3|\mathcal{T}_c |p_1,p_2,p_3\rangle 
+\langle q_1,q_2,q_3|\mathcal{T}_d| p_1,p_2,p_3\rangle\\
&=
\frac{1}{3!}\sum_{n=1}^3\sum_{m=1}^3\,
v(q_{\bar{n}},q_{\barr{n}})
\underbrace{
\Bigg(
\tau(\sigma(q_n))\,
T(q_n,p_m;s)\,
\tau(\sigma(p_m))\,
-2E({\boldsymbol q}_n)\tau(\sigma(q_n))(2\pi)^3\delta^3(\boldsymbol{q}_n-\boldsymbol{p}_m)
\Bigg)}_{=:\hat T(q_n,p_m;s)}
v(p_{\bar{m}},p_{\barr{m}}) \,,\nonumber
\end{align}
\end{widetext}
where $P$ is the total four-momentum of the system, $s=W^2=P^2$ and $E({\boldsymbol p})=\sqrt{{\boldsymbol p}^2+M^2}$. All four-momenta $p_1,q_1,...$ are on-mass-shell, and the square of the invariant mass of the isobar reads $\sigma(q):=(P-q)^2=s+M^2-2WE(\boldsymbol q)$ for the spectator momentum $q$. We work in the total center-of-mass frame where ${\boldsymbol P}=\boldsymbol{0}$. The dissociation vertex $v(p,q)$ of the isobar decaying in asymptotically stable particles, e.g., $\rho\,(p+q)\to \pi(p)\pi(q)$, is chosen to be cut-free in the relevant energy region, which is always possible. The notation is such that, e.g., for a spectator momentum $q_n$ the isobar decays into two particles with momenta $q_{\bar{n}}$ and $q_{\barr{n}}$.

For the present study we choose the dissociation vertex to be of a particularly simple form, $v(p,q):=\lambda f((p-q)^2)$ with $f$ such that it is 1 for $(p-q)^2=0$ and decreasing sufficiently fast for large momentum difference, e.g.,  $f(Q^2)=\beta^2/(\beta^2+Q^2)$ to regularize integrals of the scattering equation. Note that one is by no means obliged to use form factors but can instead formulate the dispersive amplitude through multiple subtractions rendering it automatically convergent, see Eq.~(14) in Ref.~\cite{Mai:2017vot}.

Imposing three-body unitarity and a general ansatz for the isobar-spectator scattering amplitude $T$ in Eq.~\eqref{eq:t33full} one obtains
\begin{align}\label{eq:T}
T(q,p;s)= 
B(q,p;s)-
\int
\frac{\mathrm{d}^3\boldsymbol{l}}{(2\pi)^3}
B(q,l;s)
\frac{\tau(\sigma(l))}{2E({\boldsymbol l})}
T(l,p;s)
\end{align}
with
\begin{align}\label{eq:B}
B(q,p;s)
=
&\frac{-\lambda^2f((P-q-2p)^2)f((P-2q-p)^2)}
{2E({\boldsymbol q}+{\boldsymbol p})\left(W-E({\boldsymbol q})-E({\boldsymbol p})-E({\boldsymbol q}+{\boldsymbol p})+i\epsilon\right)}\non
&+C(q,p;s)\,,
\end{align}
where $p$ and $q$ denote the on-shell four-momenta of the in- and outgoing spectator, respectively. Additional terms $C$ that are real functions of energy $W$ and momenta in the physical region as demanded by three-body unitarity (three-body forces) can be added to $B$, see discussion in Ref.~\cite{Mai:2017vot}. We postpone the introduction of multiple isobars and of spin and isospin for the isobars and the stable particles to future work. As demonstrated in Ref.~\cite{Mai:2017vot} the algebraic form of the isobar propagator is fixed up to regular terms and can be written as
\begin{align}\label{eq:tau}
\frac{1}{\tau(\sigma(l))}=
&\sigma(l)-M_0^2\\
&-
\int \frac{d^3\boldsymbol{k}}{(2\pi)^3} 
\frac{\lambda^2 (f(4\boldsymbol{k}^2))^2}{2E({\boldsymbol k})(\sigma(l)-4E({\boldsymbol k})^2+i\epsilon)} \,,\nonumber
\end{align}
where $M_0$ is a free parameter that can be used to fit (together with $\lambda$ and $\beta$) the two-body amplitude corresponding to the considered isobar, which can be  defined symbolically via $T_{22}:=v\tau v$. We will refer to the integral term in Eq.~(\ref{eq:tau}) as self-energy in the following.

One possible strategy to solve the above system of equations in the infinite volume is to reduce the three-dimensional integral equation to one-dimensional ones via partial-wave projection. To demonstrate this, we simplify henceforth the notation as $\hat T(\boldsymbol{q},\boldsymbol{p};s)$ etc., since all particles in the above equations are on their mass shell, and write the decomposition as
\begin{align}
\label{tpw}
&\hat T({\boldsymbol q}, {\boldsymbol p};s)=4\pi\sum_{l,m}\sum_{l',m'}\,Y_{l m}(\hat {\boldsymbol q})\, \hat T_{l m,l'm'}(q,p;s)\,Y^*_{l' m'}(\hat {\boldsymbol p})\,, \\
&\hat T_{l m, l'm'}(q,p;s)=\frac{1}{4\pi}
\int d\Omega_{\boldsymbol q}\int d\Omega_{\boldsymbol p}\
Y^*_{lm}(\hat{\boldsymbol q}) \hat T({\boldsymbol q}, {\boldsymbol p};s)Y_{l'm'}(\hat{\boldsymbol p}) \,.\nonumber
\text{~~~}
\end{align}
Here, $q:=|\boldsymbol{q}|$ and $p:=|\boldsymbol{p}|$ (not to be confused with the four-vector notation used before), and $Y_{l m}$ are the usual spherical harmonics. The kernel $B({\boldsymbol q},{\boldsymbol p};s)$ is partial-wave projected analogously to Eq.~\eqref{tpw} as
\begin{align}
\label{bpw}
&B({\boldsymbol q}, {\boldsymbol p};s)=4\pi\sum_{l,m}\sum_{ l' m'}\,Y_{l m}(\hat {\boldsymbol q})\,B_{l}(q,p;s)\,\delta_{ll'}\delta_{mm'}\,Y^*_{l' m'}(\hat {\boldsymbol p})\,,\\
&B_{l}(q,p;s)=\frac{1}{4\pi}
\int d\Omega_{\boldsymbol q} \int d\Omega_{\boldsymbol p}
Y^*_{lm}(\hat{\boldsymbol q})B({\boldsymbol q}, {\boldsymbol p};s)Y_{lm}(\hat{\boldsymbol p}) \,.\nonumber
\end{align}
After partial-wave projection the integral equation~\eqref{eq:T} decouples into a set of one-dimensional equations that can be solved at a given value of $s$ by rotating the integration contour into the complex plane to avoid the cut due to the partial-wave projection of the kernel $B_l$. See, e.g., Ref.~\cite{Doring:2009yv} for this standard procedure.

\section{Finite-Volume Amplitude}
\label{sec:finvol}

\begin{figure}
\begin{center}
\includegraphics[width=0.99\linewidth,trim=0.5cm 0.75cm 0cm 0cm]{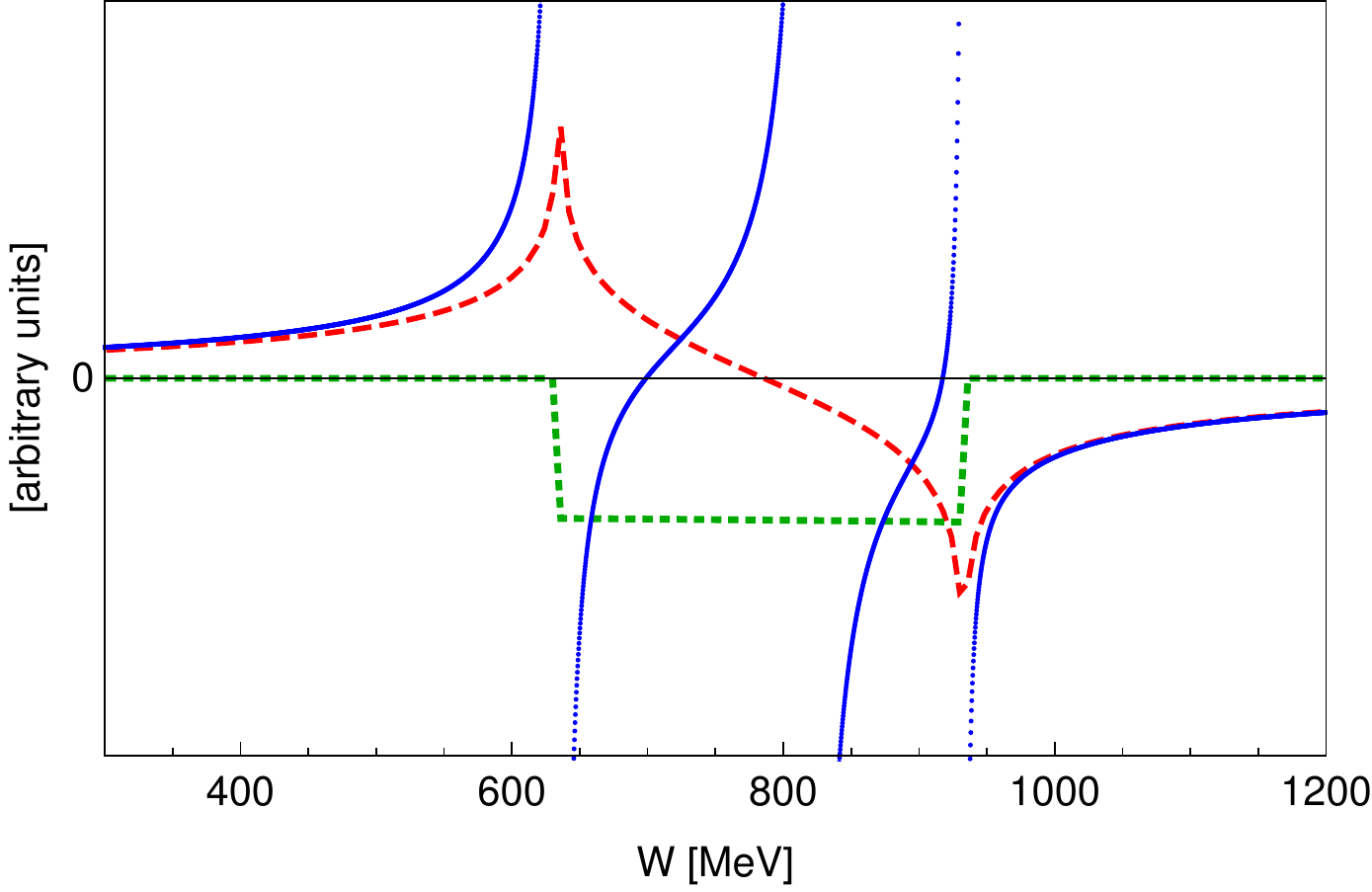}
\end{center}
\caption{Red dashed line (green dotted line): Real (imaginary) part of the infinite-volume $S$-wave projection of the potential $B_0$ from Eq.~(\ref{bpw}) with $M=138$~MeV, $\lambda=3476$ and $\beta=5000$~MeV. In comparison, the finite-volume projection $B_{11}^{A_1^+}$ from Eq.~(\ref{eq:bfinvol}) for the transition from shell 1 to shell 1 at $L=6$~fm (blue dots).}
\label{fig:compapwa}
\end{figure}

The formalism presented in the last section has to be modified when scattering in the finite volume is considered. In the following, we discuss such modifications and possible caveats. In the finite cubic volume with periodic boundary conditions the momenta are discretized. In particular, in a cube of side length $L$ only the following three-momenta are allowed (organized by ``shells'')
\begin{align}
\label{eq:discrete}
{\boldsymbol q}_{ni}&=\frac{2\pi}{L}\,{\boldsymbol r}_i 
\text{~for~} 
\{{\boldsymbol r}_i\in \mathds{Z}^3 | {\boldsymbol r}_i^2=n,i=1,\dots,\vartheta(n)\} \ .
\end{align}
Here, $\vartheta(n)=1,6,12,\dots$ indicates the multiplicity (number of points in shell  $n=0,1,2,\dots$) that can be calculated as described, e.g., in Ref.~\cite{Doring:2011ip}. Note that for each shell $n$ there is only a finite number of points, leading to the fact that the complete set of basis functions in a given shell is also finite. The partial-wave projection in finite volume can be understood in terms of the irreducible representations of cubic symmetry, see, e.g., Ref.~\cite{Bernard:2008ax}. For regular potentials, the regular summation theorem guarantees that one can proceed with the infinite-volume partial-wave projection, see, e.g., Ref.~\cite{Doring:2012eu}. However, three-body unitarity necessarily induces a potential that is singular at energies above the three-body threshold, i.e. when $W>3m$ (see Eq.~(\ref{eq:B})). While the general techniques for an expansion of a function on a shell will be discussed elsewhere~\cite{inprep}, we restrict ourselves here to the $A_1^+$ representation that in the infinite volume corresponds to $S$-wave, $G$-wave, \dots scattering. The integration over the solid angle breaks down to a summation over the points in a shell as
\begin{align} 
\int d\Omega_{\boldsymbol p_n}\to \frac{4\pi}{\vartheta(n)}\sum_{i=1}^{\vartheta(n)} \,.
\end{align}
Using the same normalization as introduced in Eqs.~(\ref{tpw}, \ref{bpw}), this replacement leads to a projection of 
$F(\boldsymbol{q}_{ni},\boldsymbol{p}_{mj};s)\in\{\hat T(\boldsymbol{q}_{ni},\boldsymbol{p}_{mj};s),\, B(\boldsymbol{q}_{ni},\boldsymbol{p}_{mj};s)\}$ to $A_1^+$ of in- and outgoing discretized momenta on the fixed shells $m$ and $n$, respectively,
\begin{align}\label{eq:A+proj}
&F^{A_1^+}_{nm}(s)=\\
&
~~~~\frac{4\pi}{\vartheta(n)\vartheta(m)}
\sum_{i=1}^{\vartheta(n)}\sum_{j=1}^{\vartheta(m)}
\chi^{A_1^+}(\boldsymbol{\hat q}_{ni})
F(\boldsymbol{q}_{ni},\boldsymbol{p}_{mj};s) 
\chi^{A_1^+}(\boldsymbol{\hat p}_{mj})\,,\nonumber
\end{align}
where $\chi^{A_1^+}$ denotes the real-valued, normalized  
${(\frac{4\pi}{\vartheta(n)}\sum_{i=1}^{\vartheta(n)}\chi^{A_1^+}(\boldsymbol{\hat q}_{ni})\chi^{A_1^+}(\boldsymbol{\hat q}_{ni})=1)}$ basis vector associated with $A_1^+$.
Since $\chi^{A_1^+}$ is orthogonal to the basis-vectors associated with other irreducible representations, the projection $B^{A_1^+}_{mn}(s)$ decouples and the finite-volume scattering problem can be solved individually in each representation. The basis vectors can be defined as linear combinations of cubic harmonics and their form and degeneracy change from shell to shell~\cite{inprep}. For $n\leq 8$ the shell index can be neglected, i.e. for all shells $n\leq 8$ the basis vectors read $\chi^{A_1^+}=Y_{00}=1/\sqrt{4\pi}$. For simplicity we restrict ourselves to this case and define ${set_8:=\{0,1,2,3,4,5,6,8\}}$ as the set of indices of non-empty shells.
This effectively introduces a cutoff for the problem at hand, which for instance lies around $\Lambda\sim1$~GeV for a typical volume of $L=3$~fm. 


According to Eq.~\eqref{eq:B} and \eqref{eq:A+proj} the projection of the driving term to $A_1^+$ reads
\begin{widetext}
\begin{align}\label{eq:bfinvol}
B_{nm}^{A^+_1}(s)
=
-\frac{1}{\vartheta(n)\vartheta(m)}
\sum_{i=1}^{\vartheta(m)}\sum_{j=1}^{\vartheta(n)}&\Bigg(
\frac{\lambda^2f((W-2E_m-E_n)^2-|2\boldsymbol{p}_{mi}+\boldsymbol{q}_{nj}|^2)
f((W-2E_n-E_m)^2-|2\boldsymbol q_{nj}+\boldsymbol{p}_{mi}|^2)}
{2E(\boldsymbol q_{nj}+\boldsymbol{p}_{mi})
(\sqrt{s}-E_m-E_n-E(\boldsymbol{q}_{nj}+\boldsymbol{p}_{mi}))}~~~~~~~~~~\non
&\hspace{9cm}
+C(\boldsymbol{q}_{nj},\boldsymbol{p}_{mi};s)\Bigg) \ .
\end{align}
\end{widetext}
Note that the partial-wave projected singular potential $B^{A_1^+}$ has an entirely different structure than its infinite-volume counterpart $B_0$. As Eq.~(\ref{bpw}) shows, for $W>3m$ the simple pole in $B$ induces logarithmic branch points and imaginary parts in $B_0$ while the projection of Eq.~(\ref{eq:bfinvol}) induces a finite number of simple poles depending on the shell indices $m$ and $n$.
For $m=n=1$ the resulting finite-volume projection to $A_1^+$ and the infinite-volume projection to $S$-wave using Eq.~(\ref{eq:A+proj}) and (\ref{bpw}), respectively, are shown in Fig.~\ref{fig:compapwa}.
Clearly, the infinite-volume projection is a good approximation as long as the potential is regular. Once it becomes singular the infinite-volume partial wave develops an imaginary part and logarithmic branch points while the finite-volume counterpart remains real but develops poles. The first pole occurs at the non-interacting energy at which the incoming and outgoing spectator momenta are back-to-back $(\theta=\pi)$, the second pole corresponds to a scattering angle of $\theta=\pi/2$ and the third pole to $\theta=0$. This exhausts the possibilities for $m=n=1$. Higher shells exhibit more poles.


\begin{figure}
\begin{center}
\includegraphics[width=0.99\linewidth,trim=0.5cm 0.5cm 0cm 0cm]{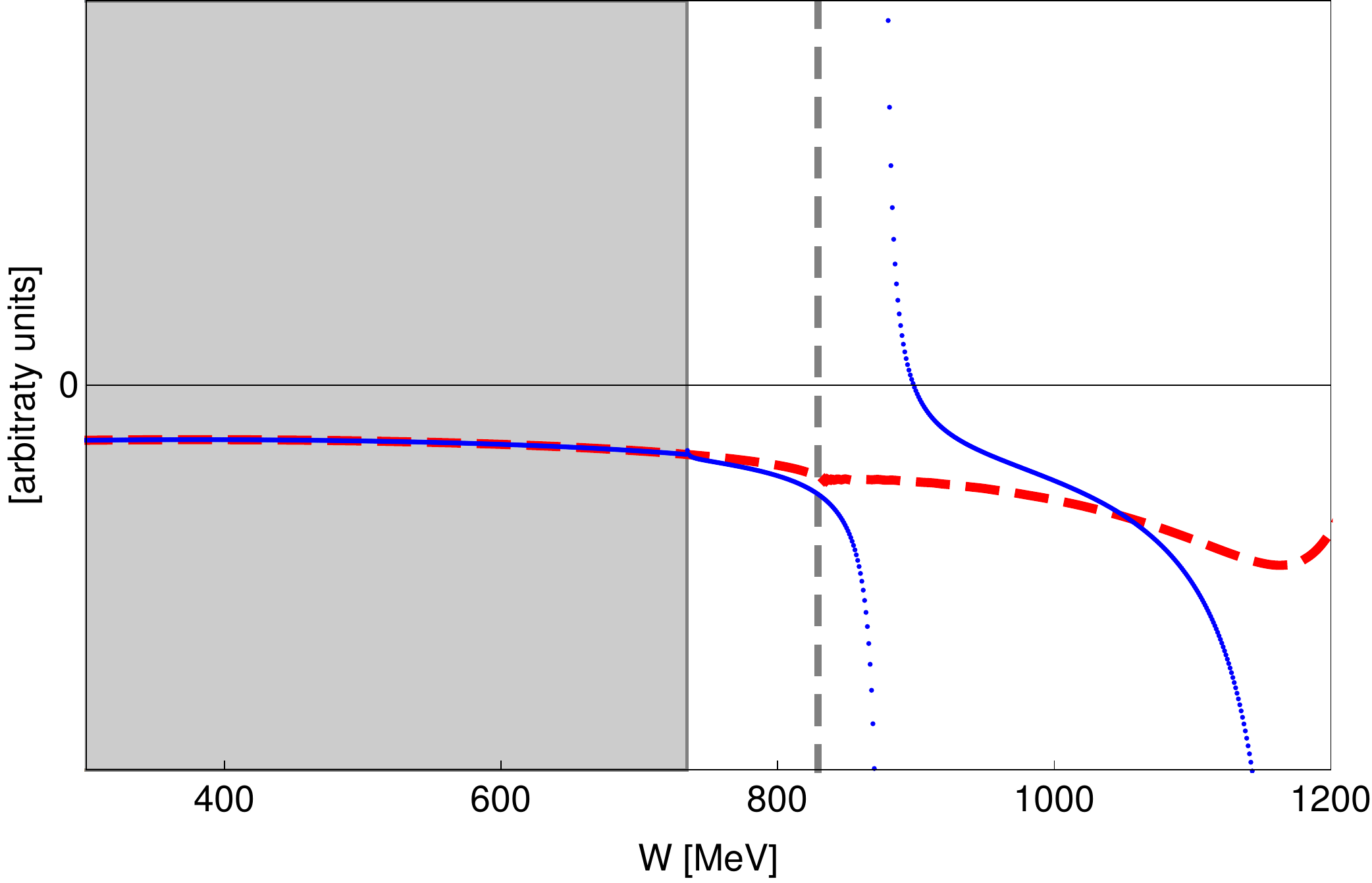}
\end{center}
\caption{Isobar propagator $\tau$ in the finite volume (blue dots) and the real part of the infinite volume one (red dashed line) for a given boost $\boldsymbol{l}=(0,0,2\pi/L)$ and $L=3.5$~fm, $\lambda=3476$, $\beta=5000$~MeV, $M_0=847$~MeV. The gray dashed line denotes the onshell-condition of two particles ($\sigma(\boldsymbol{l})=(2M)^2$), whereas the gray area represents the energy range for which $\sigma(\boldsymbol{l})\le0$.}
\label{fig:tau}
\end{figure}

For the isobar propagator in Eq.~(\ref{eq:tau}) one has to keep in mind that the self energy is evaluated in the isobar center-of-mass frame. In the finite volume, however, the allowed momenta given by Eq.~(\ref{eq:discrete}) are defined in the three-body rest frame at $\boldsymbol{P}=\boldsymbol{0}$. For the calculation of the finite-volume self-energy defined in the following one therefore has to boost the momenta to the isobar rest frame according to
\begin{widetext}
\begin{align}\label{eq:boost}
\boldsymbol{k}\to\boldsymbol{k}^*(\boldsymbol{k},\boldsymbol{l})
:=
\boldsymbol{k}+
\boldsymbol{l}\Bigg(\frac{\boldsymbol{k}\cdot \boldsymbol{l}}{\boldsymbol{l}^2}\Big(\frac{\sqrt{\sigma(\boldsymbol{l})}}{W-E(\boldsymbol{l})}-1\Big)+
\frac{\sqrt{\sigma(\boldsymbol{l})}}{2(W-E(\boldsymbol{l}))}\Bigg)
\text{~~with~~}
J(\boldsymbol{l})=\frac{\sqrt{\sigma(\boldsymbol{l})}}{W-E(\boldsymbol{l})}
\end{align}
denoting the corresponding Jacobian. The isobar propagator in the finite volume then takes the form
\begin{align}\label{eq:finvolTAU}
\tau_{m}^{-1}(s)
&=
\sigma(\boldsymbol{l}_{mi})-M_0^2
-\frac{1}{L^3}
\sum_{\boldsymbol{x}\in \mathds{Z}^3}
\frac{J(\boldsymbol{l}_{mi})\,
\left(\lambda f\left(4\boldsymbol{k}^*(\frac{2\pi}{L}\boldsymbol{x},\boldsymbol{l}_{mi}\right)\right)^2}
{2E\left(\boldsymbol{k}^*\left(\frac{2\pi}{L}\boldsymbol{x},\boldsymbol{l}_{mi}\right)\right)
\left(\sigma(\boldsymbol{l}_{mi})-4\left(E(\boldsymbol{k}^*(\frac{2\pi}{L}\boldsymbol{x},\boldsymbol{l}_{mi}))\right)^2\right)} \,.
\end{align}
\end{widetext}
Note that for shells $m\leq 8$ the isobar propagator $\tau$ depends only on the shell index and not the actual point coordinates ${\boldsymbol l}_{mi}$ within that shell (just like in the infinite volume). The reason is that in these (lower) shells there is only one way to express $\boldsymbol{l}_{mi}^2$ as sum of three squares and the sum in the self-energy part runs over $\boldsymbol{x}\in \mathds{Z}^3$. However, for the ninth shell one has $9=(\pm 3)^2+0^2+0^2=(\pm 1)^2+(\pm 2)^2+(\pm 2)^2$. One could distinguish such degeneracies for $n\ge 9$ with an additional index for $\tau_m$ and proceed as before but here we restrain from this obvious generalization for simplicity. For a given absolute value of the spectator momentum $\boldsymbol{l}_{mi}$, the range of validity of the boost formula~(\ref{eq:boost}) and, therefore, for the discretized propagator $\tau$ is limited to $\sigma({\boldsymbol l}_{mi})>0$. However, already for the two-particle threshold $\sigma({\boldsymbol l_{mi}})<(2M)^2$ the regular summation theorem applies and the sum can be replaced by the integral, see Eq.~\eqref{eq:T}, up to exponentially suppressed terms. Note that the momentum cutoff $\Lambda=2\pi\sqrt{8}/L$ implies an upper bound on the total energy $W$. For $W>W_{\rm max}$, where $W_{\rm max}$ is given by $\Lambda=\lambda^{1/2}(W_{\rm max}^2,M^2,4M^2)/(2W_{\rm max})$, not all finite-volume poles of $\tau_m(W^2)$ are taken into account. This would violate the three-body quantization condition. The corresponding form of the considered isobar propagator on the shell $n=1$ is depicted in Fig.~\ref{fig:tau}.

Finally, we replace three-dimensional integrations in Eqs.~\eqref{eq:T} and \eqref{eq:tau} by the summations over the discretized lattice momenta via 
\begin{align}\label{eq:integraltosum}
\int\frac{d^3{\bf q}}{(2\pi)^3}
\to
\frac{1}{L^3}
\sum_{n\in set_8}
\sum_{i=1}^{\vartheta(n)}\,.
\end{align}
Thus, the system of equations to solve reads 
\begin{align}\label{eq:finvolThat} 
\hat T_{nm}^{A^+_1}(s)&
=\tau_n(s)T^{A_1^+}_{nm}(s)\tau_m(s)
-2E_n\tau_n(s)\frac{L^3}{\vartheta(n)}\delta_{nm} \ ,\\
\label{eq:finvolT}
T_{nm}^{A^+_1}(s)
&=
B_{nm}^{A^+_1}(s)
-\frac{1}{L^3}\sum_{x\in set_8}\vartheta(x)
B_{nx}^{A^+_1}(s)
\frac{\tau_{x}(s)}{2E_x}
T_{xm}^{A^+_1}(s)\,,
\end{align}
where $E_n=\sqrt{M^2+(2\pi/L)^2 n}$. 
Finally, re-ordering Eqs.~(\ref{eq:finvolThat}, \ref{eq:finvolT}) we obtain 
\begin{align}\label{eq:finvolThat-short}
\hat T^{A_1^+}(&s)=\Big[ X(s)B^{A_1^+}(s)X(s)+X(s)\tau(s)^{-1}\Big]^{-1}\\
&\text{~~for~~}
X(s):={\rm Diag}_{n\in set_8}\Big(\frac{\vartheta(n)}{2E_n(s)L^3}\Big)\,,\nonumber
\end{align}
and all other elements being (in case of $\tau(s)$, diagonal) matrices in the $8\times8$ space w.r.t the first eight shells. Obviously, $\hat T^{A_1^+}(s)$ is singular, iff
\begin{align}\label{eq:determinant}
{\rm Det}[B^{A_1^+}(s)X(s)+\tau(s)^{-1}]=0\,.  
\end{align}
This quantization condition represents the final and central result of the present work. It determines the positions of the energy eigenvalues for a given set of free parameters $\lambda, \beta, M_0$ in the finite volume. In the next section we will use this result to study several technical issues and solution strategies.

\section{Determination of finite volume energy eigenvalues}\label{sec:numeric}

\subsection{General strategy}

The set of equations (\ref{eq:bfinvol}) through (\ref{eq:finvolTAU}) defines the finite-volume $\hat T$ matrix with the isobar-spectator interaction projected to the $A_1^+$ irreducible representation. The real pole positions of $\hat T(s)$ correspond to the energy eigenvalues as determined on the lattice. In an actual analysis of such eigenvalues one would tune the free parameters of the amplitude until the pole positions fit them. A strategy would take the following workflow that is related to the proposal of Ref.~\cite{Hammer:2017kms}:
\begin{enumerate}
\item \label{step1}
If possible, determine first the two-body finite-volume amplitude $T_{22}$. As we choose the particularly convenient isobar notation $T_{22}=v\tau v$, this would amount to fitting $\lambda,\,\beta,\,M_0$. However, any other parametrization of $2\to 2$ scattering would be equally valid as long as two-body unitarity in the infinite volume is respected. It is plausible that eigenvalues corresponding to the $2\to 2$ problem, potentially in moving frames, are available from the same lattice configuration on which the 3-body eigenvalues are determined. For example, the eigenvalues corresponding to $\pi\rho$ scattering in $S$-wave of Ref.~\cite{Lang:2014tia} have been determined on the same lattice configuration as for the $\pi\pi$ $P$-wave scattering of Ref.~\cite{Lang:2011mn}.
\item
For a finite number of 3-body eigenvalues available from a given lattice QCD simulation, determine the three-body force $C$ (cf. Eq.~(\ref{eq:bfinvol})) by fitting. As discussed before and in Ref.~\cite{Mai:2017vot}, $C$ is a real, energy and momentum-dependent function with exponentially suppressed finite-volume effects that can be added to the interaction $B$. Depending on the quality and number of data $C$ may be parametrized by more or less parameters~\cite{Landay:2016cjw}.
\end{enumerate}

With all parameters fixed, the infinite-volume amplitude can be evaluated. Through standard techniques like contour deformation in the momentum integration~\cite{Doring:2009yv} one can evaluate $T$ at complex energies $W$ to determine the pole position of resonances and their complex coupling constant, which in general is still a function of the incoming and outgoing isobar invariant masses.

It is also worth mentioning that most three-body systems require a coupled-channel treatment. The number of possible combinations of quantum numbers is in general large and the isobar-spectator formalism is designed to systematically enlarge the possible set of isobars depending on the abundance and accuracy of available lattice data.  Step \ref{step1} should then be repeated according to the chosen isobar quantum numbers. That choice can be guided by model selection techniques to determine the minimally necessary set of partial waves~\cite{Landay:2016cjw}.

\subsection{Cancellation of finite-volume divergences}

On one side, singularities in the finite-volume amplitude $\hat T$ of Eq.~(\ref{eq:finvolThat}) arise from the projected driving term $B$ in Eq.~(\ref{eq:bfinvol}) and from the three-body propagator $\tau$ of Eq.~\eqref{eq:finvolTAU}. These singularities are manifestly present in the respective expressions. For $\tau$ the situation is particularly complicated because it appears already in the re-scattering series for $T$ defined in Eq.~(\ref{eq:finvolT}) but also in the incoming/outgoing isobars of the connected amplitude (first term in Eq.~(\ref{eq:finvolThat})) and even in the disconnected topology (second term in Eq.~(\ref{eq:finvolThat})). On the other side, there are poles from genuine three-body dynamics originating from the infinite resummation implied in Eq.~(\ref{eq:finvolT}). We call them ``genuine singularities'' in the following. 

In principle, all singularities of the $3\to 3$ amplitude $\hat T$ of Eq.~(\ref{eq:finvolThat}) correspond to measurable energy eigenvalues. In particular, measurements on the lattice cannot distinguish between the connected and disconnected topologies (on the hadronic level), and cannot isolate the isobar-spectator amplitude $T$. This could be particularly problematic because, e.g., $\tau$ contains the full tower of 2-body energy eigenvalues which, depending at which external momenta $\hat T$ is evaluated, could be even boosted depending on the incoming/outgoing spectator momentum. These ``external'' singularities would severely obscure the signal from the genuine three-body dynamics. It is therefore necessary to trace potential cancellations of singularities. Note, that some of the cancellations of divergences have already been mentioned in Refs.~\cite{Polejaeva:2012ut, Briceno:2012rv}.

In the present formulation, we can pinpoint three different cancellation mechanisms very clearly. \underline{First}, we observe a multiplicative cancellation of singularities of $B^{A_1^+}$ by the zeros of $\tau(s)$ in the second term of Eq.~\eqref{eq:finvolT}. Specifically, on any shells $n$ and $m$ the kernel $B_{nm}^{A_1^+}(s)$ is singular at ${W^+=E_m+E_n+E(\boldsymbol{q}_{nj}+\boldsymbol{p}_{mi})}$ for $\boldsymbol{q}_{nj}~(j=1,..,\vartheta(n))$ and $\boldsymbol{p}_{mi}~(i=1,..,\vartheta(m))$. At the same time, the self-energy part of $\tau_m^{-1}$ has  singularities at $W^{\pm\pm}=E_m\pm E((2\pi/L)\boldsymbol{x}) \pm E((2\pi/L)\boldsymbol{x}+\boldsymbol{p}_{mi})$ summing over all possible ${\boldsymbol x}\in \mathds{Z}^3 $. Therefore, $\tau_m^{-1}$ has singularities at the same and more energies as the projected kernel $B^{A_1^+}$ so that  all singularities of $B^{A_1^+}$ cancel. \underline{Second}, the singularities of the first term of Eq.~\eqref{eq:finvolT} are still present in the solution of $T_{mn}^{A_1^+}(s)$ for all $m$ and $n$. However, they are exactly canceled as described before, considering the full matrix element $\tau_n(s)T_{nm}^{A_1^+}(s)\tau_m(s)$ in the first term of Eq.~\eqref{eq:finvolThat}. \underline{Third}, the same term has singularities stemming from the isobar propagator $\tau_n(s)$. These are exactly canceled by the contribution stemming from the disconnected diagram, i.e. the second term of Eq.~\eqref{eq:finvolThat}. 

\subsection{Quantitative example}

\begin{figure}
\begin{center}
\includegraphics[width=0.99\linewidth,trim=0.5cm 1cm 0cm 0cm]{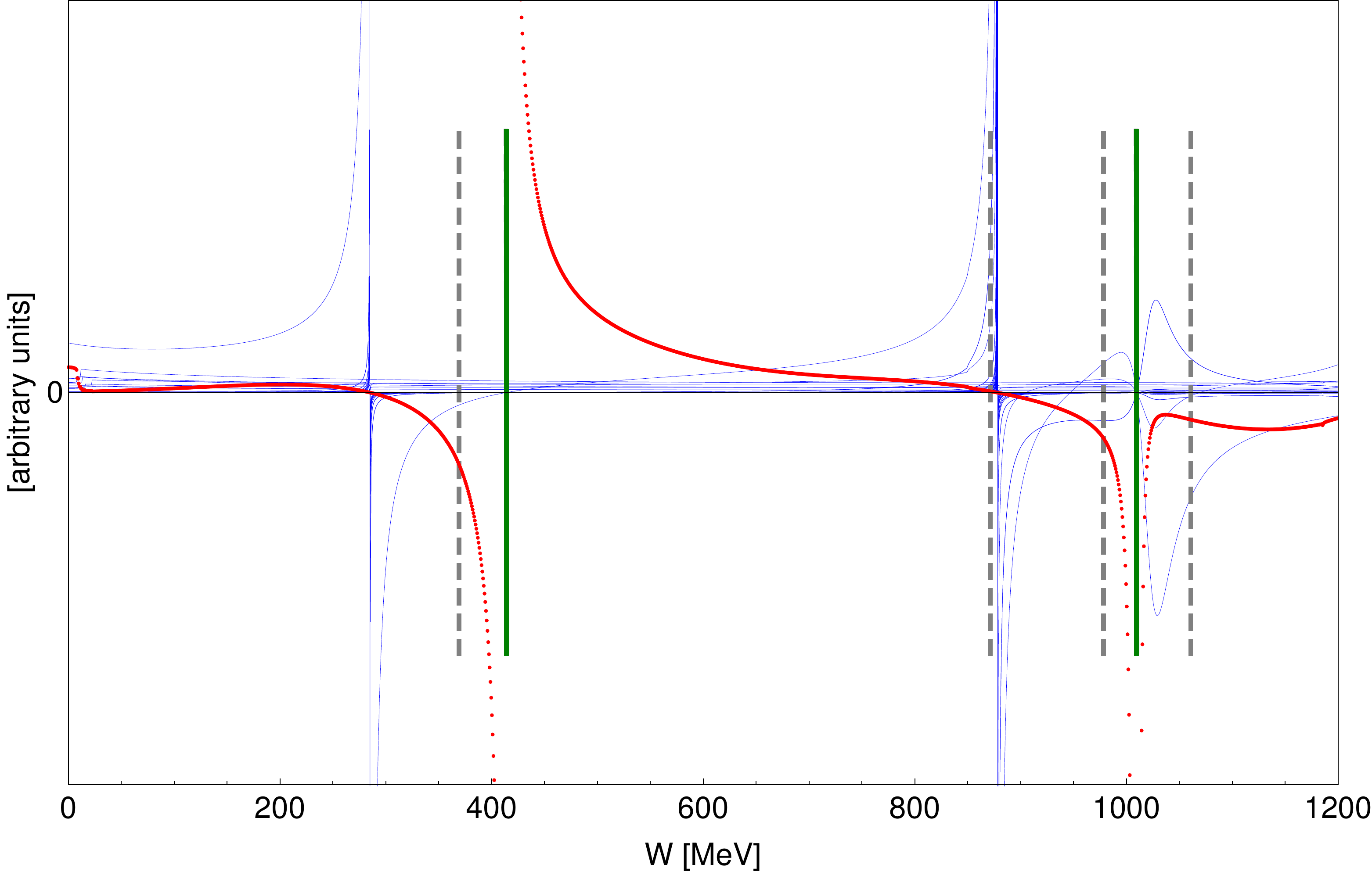}
\end{center}
\caption{The blue lines show all finite-volume  matrix elements $\hat T_{mn}^{A_1^+}$  $ (m,n\in set_8)$ for $L=3.0$~fm and $M=138$~MeV. The red dots show the left hand side of Eq.~\eqref{eq:determinant} (multiplied by a regular, non-vanishing function of energy for convenience). The dashed vertical lines show the positions of singularities of $\tau$, the solid green vertical lines those of $B^{A_1^+}$. Note that all $\hat T_{mn}^{A_1^+}$ are regular at these singularities showing that the cancellation of spurious singularities is complete.
}
\label{fig:final1}
\end{figure}

To demonstrate the quantitative implications of the quantization condition~\eqref{eq:determinant} we fix the free parameters to $\lambda=3476,\beta=5000$~MeV, $M_0=847$~MeV, such that the phase shift of the corresponding $2\to2$ amplitude is of natural size. Furthermore, the size of the lattice is chosen to be $L=3.0$~fm and the particle mass $M=138$~MeV. A more realistic case corresponding to some actual physical system will be discussed in future work. Here we are concerned with more generic features of the $3\to3$ amplitude in finite volume. 

The result is presented in Fig.~\ref{fig:final1}. We observe clear cancellations of poles arising solely from the driving term $B^{A_1^+}$ (green vertical lines) in the isobar-spectator amplitude as well as those of the isobar-propagator $\tau$ (gray dashed vertical lines). 
All finite-volume energy eigenvalues appears to be well separated from the singularities of $B^{A_1^+}$ and $\tau$, which demonstrates the numerical efficiency and applicability of the proposed framework.


\section{Conclusions}

A relativistic method for the infinite-volume extrapolation of on-shell three-body systems is proposed, based on a recently derived manifestly two and three-body unitary infinite-volume amplitude. This amplitude depends only on on-shell two-body unitary sub-amplitudes and real-valued three-body forces. Without loss of generality, we choose here an isobar parametrization for the two-body amplitude.
Three- and two-body unitarity dictate the imaginary parts of the amplitude in the infinite volume and the leading power-law effects of the finite volume when all three particles are on-shell, implying the correct three-body quantization condition. All spurious finite-volume singularities tied to the necessarily singular interaction, the two-body spectrum, and the disconnected topology cancel, leaving only singularities from genuine three-body dynamics.

The organization of lattice points in shells allows for a particularly simple parametrization of the finite-volume amplitude even after projection to the irreducible representations. A two-step workflow of first fitting internal amplitude parameters to eigenvalues from lattice QCD simulations and then evaluating the infinite-volume counterpart is proposed. A first numerical implementation for scalar-isoscalar particles is performed.

\acknowledgements
The authors thank A. Alexandru, R. Briceño, H.-W. Hammer, M. Hansen, L. Leskovec, J. Y. Pang and A. Rusetsky for useful discussions. 
This work is supported by the National Science Foundation (CAREER grant
PHY-1452055, NSF/PIF grant No. 1415459) and by the U.S. Department of Energy, Office of Science, Office of Nuclear Physics under contract DE-AC05-06OR23177.
M.D. is also supported by the U.S. Department of Energy, Office of Science,
Office of Nuclear Physics under grant No. DE-SC001658. M.M. is thankful to the German Research Foundation (DFG) for the financial support, under the fellowship MA 7156/1-1, as well as to the George Washington University for hospitality and inspiring environment.


\bibliography{finvol}

\end{document}